\documentstyle[aps,twocolumn]{revtex}

\newcommand{\beq}{\begin{equation}}
\newcommand{\eeq}{\end{equation}} 

\newcommand{\beqa}{\begin{eqnarray}}
\newcommand{\eeqa}{\end{eqnarray}}

\def\opone{\leavevmode\hbox{\small1\kern-3.8pt\normalsize1}}

\begin{document}

\title{Optical tests of quantum nonlocality: from EPR-Bell tests towards
experiments with moving observers}
\author
{N. Gisin, V. Scarani, W. Tittel and H. Zbinden\\
\protect\small\em Group of Applied Physics, University of Geneva, 
1211 Geneva 4, Switzerland}
\date{\today}

\maketitle

\begin{abstract}
Past, present and future experimental tests of quantum nonlocality are discussed.
Consequences of assuming that the state-vector collapse is a real physical
phenomenon in space-time are developed. These lead to experiments feasible with
today's technology.
\end{abstract}

\section{Introduction}\label{introduction}
Central to quantum mechanics is the superposition principle: if $\psi_1$ and
$\psi_2$ are two mathematical objects representing two possible physical
states of a quantum system, then any linear combination, in particular
$\psi_1 + \psi_2$, represents also a possible state of this system. Accordingly,
the $\psi$'s are usually called state-vectors. This
mathematical formulation of the superposition principle encompasses essentially
all peculiar aspects of quantum physics. The wave-particle dualism, for example,
is reflected by the fact that the linear addition of waves has a straightforward
meaning, but in quantum mechanics it applies even to systems that have well
localized effects, like particles. If this would be all the story, then a
pilot-wave picture \`a la De Broglie-Bohm would be most natural: the particle
is guided by a wave, like a cork guided by water waves. For example, this model provides a classical
picture of the famous two-slit experiment. But the
particle-wave dualism is not the entire story of the superposition principle. 

In this contribution, we discuss some past, present and future tests of quantum physics.
We insist mainly on two points. The first one concerns the importance of quantum
nonlocality. Indeed, it is only when the superposition principle is applied to systems
composed of a few particles that the most stringent characteristics of quantum
mechanics can be tested: if the system is not composed, then a classical picture suffice, 
as briefly mentioned above; if, on the contrary, the system is composed of many particles,
then the complexity of the situation makes it useless for experimental tests, as discussed in 
the next section. The second point we like to emphasize is the much discussed 
{\it collapse of the state vector}. Whether this collapse is a real physical phenomenon
or not is a long debate. We approach this delicate question by concentrating on
assumptions formulated in such a way that they can be tested with today's technology.
Our guide is the tension between quantum nonlocality and relativity:
if the collapse is a real physical phenomenon, then it is incompatibility with special
relativity and must thus lead to new measurable effects.

In the next section, the measurement problem and EPR-Bell tests are briefly
commented, with emphasize on the detection and the locality loopholes (paragraphs
\ref{DetLoophole} and \ref{LocalityExp}, respectively). 
In section \ref{SpookySpeedExp} four different assumptions about the collapse
of the state-vector as a physical phenomenon in space-time are presented and some experimental results
reviewed. Finally, a further assumption is discussed in section \ref{manyPsi}.

\section{A brief historical perspective}\label{hist}
Einstein, Podolsky and Rosen \cite{EPR}, and  Schr\"odinger \cite{Schrodinger} showed that
when the superposition principle is applied to multiparticle systems, situations even more bizarre
(i.e. counter to our classical intuition) than the two-slit experiment occur. A first example is the
infamous quantum measurement problem: the superposition principle predicts that if
result 1 is possible with the system+measurement apparatus final state represented
by $\psi_1$ and result 2 is possible with final state $\psi_2$, 
then, in principle, $\psi_1+\psi_2$ represents also a possible
state, although this is never observed. This measurement problem suffers from a serious
limitation: while quantum theory admits $\psi_1+\psi_2$ as a state, it predicts that 
the situation represented by $\psi_1+\psi_2$ is
in practice unobservable. This is due to the unavoidable interaction with the environment
which hides the correlations predicted by $\psi_1+\psi_2$. This "hiding" is called
decoherence. Whether one finds this solution satisfactory or not is "human dependent"!
Some are fascinated by the consistency of the quantum formalism. Some find it
strange that the fact that a peculiar prediction can't be tested is counted in favor
of the theory: isn't physics about testing the most peculiar predictions of our best
theories? Anyway, the fact remains that
the measurement problem can't be solved by experiments, at least not with today's
technology. But this is still not the end of the story.

Let's apply the superposition principle to a system composed of two spatially separated
objects A and B. If object A is in state $\alpha_1$ and object B in state $\beta_1$,
then the combined system is in state $\alpha_1\otimes\beta_1$. Consequently,
\beq
\alpha_1\otimes\beta_1+\alpha_2\otimes\beta_2
\label{a1b1a2b2}
\eeq
must also represent a possible
state of the composed system. Moreover, contrary to the measurement problem, states
of the form (\ref{a1b1a2b2}) can be produced in controlled environments and can be measured almost
perfectly (we shall come back to the "almost" in the next section). States of the form
(\ref{a1b1a2b2}) which can't be written as a product (by changing the basis) are called
entangled \cite{QIPIntro98,PQI00}. Entangled states have bizarre properties. First they allow remote state
preparation: by performing a measurement on the object A, object B is prepared in a
well defined state. Note that one can't decide the state in which to prepare B,
but once the measurement result on A is registered, the state of B is well determined. This
bizarre prediction can be tested: perform independent measurements on many identically
prepared objects and register the results. Then, select the cases where object B was
prepared in a specific state and check that the measurement outcomes on B for these cases
are compatible with the predicted state. But this is still not the entire story. 

Bell has showed that the situation is even more bizarre \cite{Bell64}. 
Let $a$ and $a'$ denote
possible measurements on A with results $\alpha$ and $\alpha'$, respectively, and similarly $b$ and $b'$ 
measurements on B with results $\beta$ and $\beta'$. Then $(\alpha,\beta)$ and $(\alpha,\beta')$
denote possible joint results when $a$ is measured simultaneously with $b$ and $b'$, respectively. 
Now, it seems almost obvious that
if the objects A and B are spatially separated, then results on A do not depend on what is 
measured on B. This locality assumption implies that if result $\alpha$ is registered, 
this does not depend on which
measurement $b$ or $b'$ was performed on B, and vice-versa. Hence, ascribing to the results the values $\pm1$,
they satisfy the following inequality:
\beqa
\alpha\cdot\beta+\alpha\cdot\beta'+\alpha'\cdot\beta-\alpha'\cdot\beta'&=&  \\
\alpha\cdot(\beta+\beta')+\alpha'\cdot(\beta-\beta')&\le&2
\eeqa
Consequently, the locality assumption implies that the expectation values $E(a,b)\equiv Mean\{\alpha\cdot\beta\}$
satisfy the CHSH-Bell inequality \cite{CHSHBell}:
\beq
E(a,b)+E(a,b')+E(a',b)-E(a',b')\le2
\label{CHSH}
\eeq
But, according to quantum mechanics, this is not so! All entangled pure states violate the
inequality (\ref{CHSH}) for some properly chosen measurements \cite{Gisin91}.

Consequently, quantum theory is nonlocal. But is that serious? Should we bother?
It is instructive to recall that the Einstein-Bohr debate was
considered  during tens of years as mere philosophy. Then, came John Bell and his inequality who turned the
question into experimental physics\footnote{Actually, Bell's original inequality \cite{Bell64} is untestable because it
assumes perfect correlations. It is the merit of Clauser, Horn, Shimony and Holt \cite{CHSHBell} to have derived the
inequality (\ref{CHSH}) well suited for experimental investigations.}. More recently, A. Ekert \cite{EkertQC91} 
and Ch. Bennett, G. Brassard and D. Mermin \cite{BBMQC} demonstrated that quantum correlation
can be used for secure communication, hence the question became part of applied physics!
No doubts: quantum nonlocality is central both for our understanding of the most precise theory
ever produced and for the huge potential applications of quantum technologies,
for example in the field of quantum communication and information processing
\cite{QIPIntro98,PQI00}. 

When Bell discovered his inequality, a first surprise was that no existing experimental
data could be used to test the inequality. Hence, new experiments had to be designed and
carried out. The first experiments gave conflicting results, a fact which probably explains
in part why Clauser's experiment is too often forgotten \cite{Clauser72}. 
Since Aspect  and co-workers experiments in the early 80's \cite{Aspect80s} 
many laboratories around the world performed experiments confirming the bizarre 
predictions of quantum mechanics (see \cite{Weihs98,Tittel98} and references therein). A few years
ago, we demonstrated that the experiment can also be brought out of the lab \cite{Tittel98,Tittel99}. Using
standard telecom fibers, we could violate Bell inequalities by photons that propagated
8 and 9 km, respectively, and were analysed in two Swisscom stations distant by more
than 10 km. This demonstrated the robustness of entanglement and its potential for
applications in quantum communication.

But the story does not end with all these beautiful experiments! Indeed, first
these experiments use real, i.e. non-ideal devices, hence suffer from some
loopholes. Next, once one admits the existence of quantum nonlocality, new questions
arise. The next two paragraphs, \ref{DetLoophole} and \ref{LocalityExp},
briefly summarise the detection and the locality
loopholes. Then, the sections \ref{SpookySpeedExp} and \ref{manyPsi} 
present ideas for new tests and some first results.

\subsection{The detection loophole}\label{DetLoophole}
Real detectors have a finite efficiency. Actually, almost all experiments to date used
detectors with efficiencies below 50\%, this mean that the detectors miss most of
the particles and that the correlations are measured on less than a quarter of
the particle pairs. From the standard quantum mechanics point of view, this is
easy to explain and does not cause any problem. However, if one admits that additional
variables, not yet discovered, could exist, then this is a serious problem \cite{DetLoophole}. For
example suppose that we did not know that photons have a variable called polarization.
Polarizer and polarization sensitive detectors would nevertheless exist! Hence, from the
local variable point of view it is very plausible that the detectors and other elements
we use in our experiments are sensitive to these additional variables, even if we are 
not (yet) able to master them. It is not difficult at all to device models explaining all
existing results using only local variables and the assumption that the detectors are
sensitive to these local variables (see e.g. \cite{NBGisin99}).

\subsection{Experiments closing the locality loophole}\label{LocalityExp}
In most tests of Bell inequality, the settings $a$ and $b$
are changed very slowly. Often one even keeps
them fixed until enough data are acquired to compute the corresponding correlation
$E(a,b)$. Consequently, one could imagine that the analysers somehow influence
the source and that the photon pairs are emitted in a state which depends on the
analyzers. Although such a solution would require a completely new kind of influence,
it would not require any superluminal signaling. Hence it is worth testing. Aspect
and co-workers performed the first test closing this loophole, using quasi-periodic 
acousto-optic modulators \cite{aspectswitch82}. More recently,
G. Weihs and co-workers performed an
experiment using fast polarization modulators \cite{Weihs98}. In this experiment, the settings
were chosen by random generators, based on quantum mechanics. W. Tittel and colleagues 
 also performed an experiment which can be considered closing this loophole 
\cite{Tittel99,ZbindenSwitch}. In the later 
the random choice was also of quantum origin, but implemented using an internal
degree of freedom of the tested photons, independent from the variable used to test the
inequality. 

As shown in ref. \cite{ZbindenSwitch}, these tests of the locality loophole are not independent of
the detection loophole. The story on the loopholes in Bell tests is thus not yet completely
closed, although a result contradicting quantum mechanics in these nowadays standard 
experiments would come as a huge surprise.

\section{"Speed" of the "spooky action at a distance}\label{SpookySpeedExp}
Let's assume that the collapse of the wavepacket is a real physical phenomenon.
Some physicists will immediately react that such an assumption is not necessary. At 
this point one could start to argue. The argument would probably turn around some form of the 
measurement problem, or involve a many-world interpretation \cite{FabricReality97}. 
But none of these can be tested with today's technology. Consequently, better than 
arguing, let's explore the consequences of the assumption!
Indeed, wouldn't it be nicer to argue against the concept of collapse as
a physical phenomena by designing and performing experiments, rather than
arguing that collapses might not be required to explain present day data? 

If there is a physical phenomenon in space-time, there must be a speed. 
What is the speed of the "collapse"
and how could one measure it? First about semantics. We don't have in mind the time it takes
for a collapse process to take place. Rather we think of the time it takes before distance systems
entangled with the measured one collapse themselves. Hence, we believe that "speed
of quantum information" is an appropriate name\footnote{Admittedly, there is the danger that one
(intentionally or not!) forgets the qualification "quantum" and concludes that
information propagates faster than light. There will always be ways to mistreat
terminology.}. Other possibilities would be "speed of quantum influence", or "speed of 
the spooky action", but we consider them as inappropriate to design an assumed real
phenomenon. So, let's call this hypothetical speed,
"speed of quantum information" and denote it $v_{QI}$. 
As any speed, $v_{QI}<\infty$ and its value depends on the reference frame. 
Accepting the results of Bell tests, we know that this speed must be superluminal,
$v_{QI}>c$. Hence, even the chronology depends on the reference frame. Assuming that there
is a cause -- a probabilistic cause in the form of a measurement outcome here -- and an effect there,
related either deterministically or probabilistically, the reference frame (if it exists)
must be determined either by the very condition of the experiment itself, or by 
cosmology. In this section, we review four possibilities. The three first ones have been tested
experimentally (though sometimes with additional assumptions). In the next section \ref{manyPsi} we elaborate on 
some possibly future directions.

\subsection{Laboratory frame}\label{LabFrame}
A first very natural assumption is that the reference frame is determined by the massive 
local environment of the experiment, i.e. by the laboratory frame, or, in case of our
long distance Bell experiment, by the Geneva frame. Indeed, as long as all the parts of
the experiment are at rest in the lab reference frame, it is a natural assumption (in
paragraph \ref{DetFrame} we consider what could happen if the different parts are in
relative motion). A testable consequence of this idea is that if the two measurements
on the two entangled particles take place simultaneously in the lab frame, then, however
large $v_{QI}$ might be, each particle produces a result before being informed
that the other particle also undergoes a measurement. Consequently, in case of perfect
simultaneity in the lab frame, the quantum correlation should disappear. In real tests
the measurements are never perfectly simultaneous. But a finite precision in the timing,
sets a lower bound to the speed of quantum information. In 1999 we performed an experiment
in which careful fiber length and chromatic dispersion adjustments provided a timing
accuracy of $\pm5$ ps over a distance of 10.6 km, setting the lower bound \cite{Zbinden1999,Zbinden2000}:
\beq
v_{QI}\ge \frac{2}{3}10^7c
\eeq 
where c denotes the speed of light. This is a large number indeed! It is tempting to
conclude that the laboratory (Geneva) frame is not the correct one. But one should be
careful. For centuries, the fastest measured speed was that of sound and the speed of light
was considered ill defined or infinite. However, first measurements of the
speed of light found values 6 orders of magnitude larger than the speed of sound in air\footnote{Clearly,
this is only an analogy. Contrary to quantum information, both sound and light carry energy and
classical information.}.

\subsection{Cosmic background radiation frame}\label{CBR}
The most natural singled out reference frame in cosmology is determined by the cosmic 
background radiation. In a recent paper we analyzed
our 1999 experiment in this context \cite{Scarani00}. The result is that our data set a lower bound of
\beq
v_{QI}\ge2\cdot 10^4~ c,
\eeq
again an impressive value. Improved experiments, in particular with faster data acquisition,
could still increase this bound by several orders of magnitude (or find new physics!).

\subsection{Detector frame}\label{DetFrame}
A third quite natural assumption is that the reference frame is determined by the massive 
device that triggers the collapse. This possibility is inspired by, though different
from Antoine Suarez ideas \cite{SuarezScarani97,Suarez97}. It is similar to the laboratory frame assumption discussed
in subsection \ref{LabFrame}. The most important difference is that different {\it trigger
devices} can be in relative motion and thus determine different frames. In particular one can
think of situations where two {\it trigger devices} both trigger a collapse before the other,
each in its own inertial reference frame! In such a case the concept of collapse as a causal
explanation collapses! 

Testing this assumption is difficult, but not impossible. Using the 5 ps timing accuracy
over 10 km obtained in our 1999 experiment, a relative speed of 50 m/s between the two
trigger devices suffices \cite{Zbinden1999,Zbinden2000}. More delicate is the question "what is a trigger device"? i.e.
what triggers the collapse? Again many physicists will answer: "nothing, since collapses do
not exist!". But we prefer to make testable assumptions. Probably, the assumption that
detectors are trigger devices would be generally considered as natural. But this makes the
experiment specially difficult: one would have to set the detectors, in our case liquid nitrogen
cooled Ge avalanche photodiodes, in motion with large velocities. 
This practical difficulty does of course in
no way reduce the interest of assuming that the detectors are the trigger devices and we do
hope to see an experiment along this line in the future. However, another interesting
assumption is that any
irreversible absorption acts as a trigger device\footnote{In \cite{Zbinden1999,Zbinden2000} we
argued that if detectors are assumed to be trigger devices, then absorbers would equivalently act
as trigger devices. Indeed, the essential physics of detectors is the irreversible absorption which 
takes place in the first microns of the upper semi-conductor layers. We still believe that this is
a valid reasoning, but prefer here to present the absorber case as an independent assumption, in
order to avoid confusion between different kinds of arguments.}. But then, how could one register the 
results? In \cite{Tittel98,Tittel99} we argue that this is still possible if one output port of the analyzer in
connected to an absorbing material and the other output port to a real detector; the only
requirement being that the absorption takes place well before the particle sees the detector,
as in fig. 1 where the absorber A is much closer to the interferometric analyzer than the 
detector APD3.
In this way, the particle emerges from the analyzer from both output ports, in superposition
with probability amplitudes as in standard quantum mechanics. The particle then first
encounters the absorber which triggers a collapse. Either the particle is absorbed and 
the collapse reduces the probability amplitude of the detector path down to zero, or the
particle is not absorbed and the collapse raises the probability amplitude of the detector 
path up to one. This is just standard quantum mechanics. Note that in this configuration 
the detectors do not trigger any choice, they only reveal the results of the measurements
actually performed by the absorbers. In practice, because of the detectors inefficiencies 
most results are never registered. We thus need to assume that the set of collected data 
constitute a fair sample of all events. This is an assumption independent from the hypothesis
that massive absorber determine the reference frames in which the collapse take place, but
it is not a new assumption: as discussed in section \ref{DetLoophole} all Bell tests to date must
use the fair sampling assumption.

Let us emphasize another aspect of this experiment. It is known since the very early days of
relativity that the time ordering of events may depend on the reference frame. In our
experiment, two correlated events are made to happen in such a way that each event in its
own natural reference frame happens first. Moreover, this relativistic effect could be realized
using an "almost every day" speed, 100 m/s, the speed of a Ferrari! 
This is of interest, even independently of the quantum collapse issue.

The experimental details are given in \cite{Zbinden1999,Zbinden2000}. The main conclusion is that the
quantum correlation are observed, even in this {\it before-before} configuration. This
result contradicts the assumption that absorbers determine the reference frame in which
the collapse propagates.

\subsection{massive choice device}{\label{ChoiceDevice}
A further interesting hypothesis assumes that the preferred frame is determined by the massive
device where the particle choses its path, i.e. the {\it choice device}\footnote{In 
\cite{Zbinden1999,Zbinden2000} we used the terminology of {\it choice device} also for 
detectors and absorbers, i.e. for what we call here {\it trigger device}.}. 
This idea is inspired by the pilot-wave picture
\`a la De Broglie-Bohm and has been first proposed by Suarez and Scarani \cite{SuarezScarani97}, and
developed in \cite{Suarez97}. In this picture, the quantum state vector is called a wave and is
assumed to guide a point particle. Hence, in this model, the particle's position is a variable
in addition to the usual quantum state vector (note that the position variable is not
hidden, quite on the contrary it is the only variable we access directly. In this model,
it is the state vector which is hidden!). Beam splitters and polarization analyzer are
examples of devices where the particle's position makes a choice, i.e. they are examples
of {\it choice devices}. It might be interesting to speculate on how
physics would have developed, in particular how the Einstein-Bohm controversy would 
have evolved, had De Broglie discovered earlier the model that Bohm published in the
1950's. But, even more interesting is that in this model it is very natural to assume
that the reference frames defined by the {\it choice devices} determined the chronology.
Accordingly, when two entangled photons encounter two beam splitters in relative motion such that
each photon is "analyzed" by its beam splitter before the other photon, then, again, a
before-before configuration happens and the quantum correlation should vanish. This is a
very natural extension of the pilot wave model. And it is a testable one. We are working on
realizing such an experiment, using traveling acoustic waves as moving beam splitters
\cite{Suarez00}.

\section{One state-vector per reference frame?}\label{manyPsi}
So far we have always assumed that there is a unique "state of the universe". Many
theorists have argued that a way out of the conflict between relativity and collapses as
real physical phenomena is to assume that the state vector itself depends on the reference
frame. Hence, each foliation of space-time has its own $\psi$. The different state vectors
are related such that they predict the same statistics of measurement outcomes. Along this way,
some relativistic models of collapses are developed \cite{Pearle99}. This raises however the question of the
uniqueness of data \cite{Omnes}. This problem also exists in classical mechanics, though there it 
reduces to the uniqueness of initial conditions. In quantum mechanics the problem is more 
delicate, because real choices happen all the time. With a unique state vector and real
collapses, unique data arise naturally in Newtonian relativity with an absolute time.
But the real problem is the compatibility between the uniqueness of quantum data and Einstein
relativity, where timing is relative.

Associating one state vector to each reference frame does not solve the uniqueness of data
problem. But again it is instructive to look at long distance quantum correlation with
moving devices and to formulate testable hypothesis. In figure 1 with the moving absorber (the wheel) there
are, in the context under consideration, two state vectors. The first one, $\psi_1$, is associated to the
moving absorber, the other one, $\psi_2$, is associated to all the other devices, in particular to the
detectors. Hence, all measurement results to date are compatible with the idea that the detectors
see only $\psi_2$, and not $\psi_1$, while the absorber only trigger the collapse of $\psi_1$
without affecting $\psi_2$. To test this, one should replace the moving absorber by a real
detector. The assumption is that the data from this moving detector show no correlation with
the data from the distant entangled particle, but the data from detectors at rest shows
the quantum correlation. A surprising prediction of this assumption is that in some cases both 
the moving and the static detectors detect the particle! and sometimes none detects it
(even assuming perfect detectors). On the mean, the number of counts of each detector 
corresponds to the standard predictions, but not in individual cases. There would be a serious question
of energy conservation. But is this really new? Already in the old EPR paradox, the 
momentum correlation corresponds to kinetic energy correlation: the kinetic energy measured
on one particle determines the energy of the other particle. Hence, the distant particle (i.e. not
in direct contact with the measurement apparatus) can end in states with different energies. Where
does this energy come from or go to?

\section{Conclusion}\label{concl}
The story of quantum nonlocality is fascinating. 
We could take advantage of this to promote Physics! We already learned a lot from
it, in particular about the weirdness and importance of entanglement in the quantum world.
We are also learning how important entanglement can be for our technology, namely that we
could exploit entanglement to perform tasks classically impossible, like quantum
cryptography and many others. This field is nowaday well recognized
under the general name of quantum information processing \cite{QIPIntro98,PQI00}. But the story is not at the end.
Using new technologies, revolutionary assumptions can be tested. Admittedly, some of the
assumptions discussed in this contribution are wild. But they are testable with today's
technology and they provide great intellectual excitement.
Moreover, they are very natural in the context of the debate about the existence or 
non-existence of state-vector collapses, a
question which triggered animated discussions and disputes since the very early times of quantum mechanics,
100 years ago.
It is worth realizing that there are not 
that many alternatives to the existence of "real collapses": without collapses the logical implication
of the superposition principle is that all possibilities co-exist, in the form of some multiverse
\cite{FabricReality97}. But this huge superposition can't be tested directly.
Isn't it nice to argue in favor or against the concept of collapse as
a physical phenomena by designing and performing experiments of the kind discussed in
this contribution!

\small
\section*{Acknowledgements}
This work would not have been possible without the financial support of the
"Fondation Odier de psycho-physique". It also profited from support by Swisscom and the Swiss
National Science Foundation. We would like to thank A. Suarez and Ph. Eberhart for very
stimulating discussions.

\section*{Figure caption}
Schematic of the experiments discussed in sections \ref{LabFrame}, \ref{CBR} and
\ref{DetFrame}. It consists of a photon pair source entangled in energy-time and two 
interferometric analyzers separated by 10.6 km, see \cite{Tittel99}. 
In the first experiments, the avalanched photodiodes APD1 and APD2 are set at
exactly the same distance from the source, so that in the Geneva reference frame
they register counts simultaneously with a timing accuracy of $\pm5$ ps.
In a second experiment, the detectors APD1 and APD2 are replaced by two absorbing surfaces.
The absorber A is static and the second absorber is moving at 100 m/s on a rotating wheel. They
are adjusted such as to be at the same distance from to source within $\pm1$ mm. 
With this precision, each absorber encounters a member of the photon pairs before the
other, each in its own (quasi) inertial reference frame.
The detectors APD3 and APD4 are connected with longer fibers such that
each photon meets first the absorber, next the detector. 
For further information about the experiment, see \cite{Zbinden2000}.

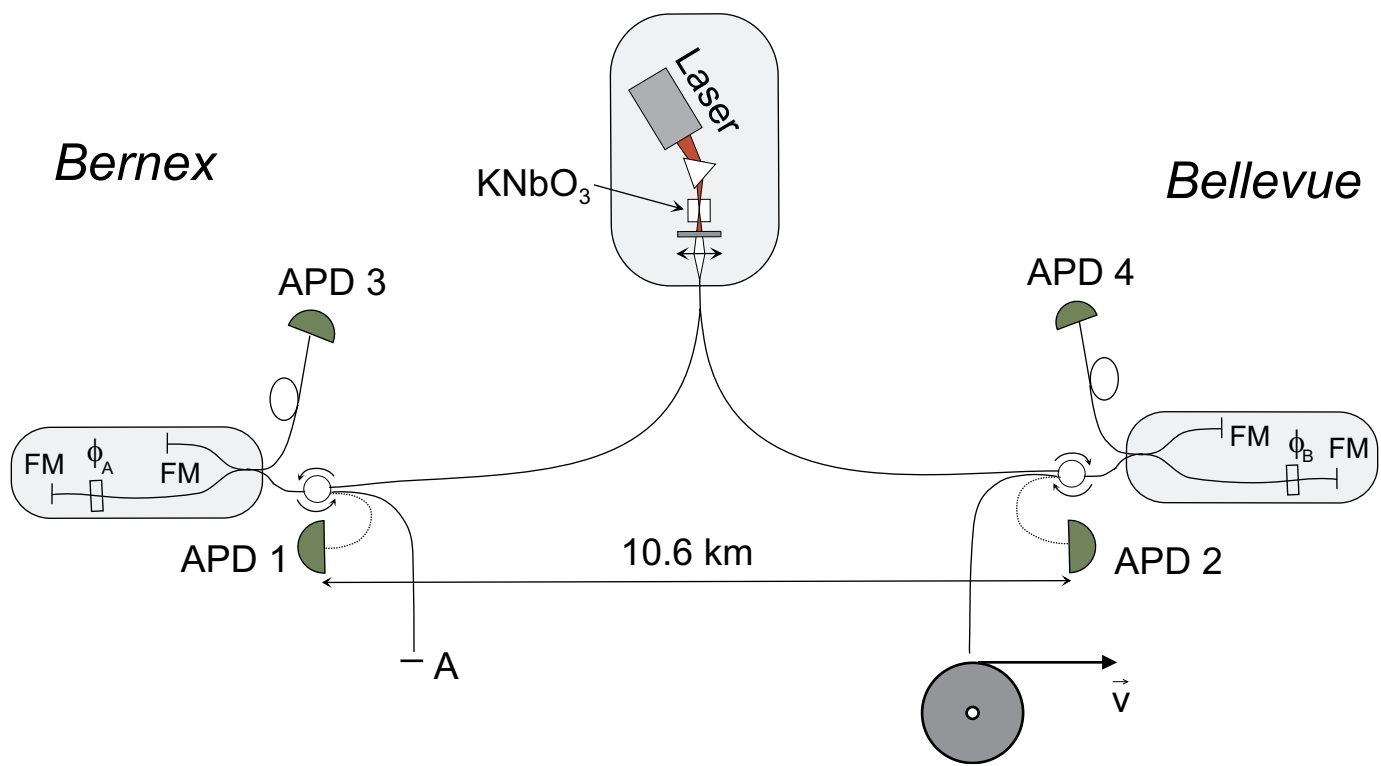
\end{document}